\documentclass[preprintnumbers]{revtex4}
\usepackage{amsfonts}
\usepackage{amsmath}
\usepackage{hyperref}
\usepackage{amssymb}
\usepackage[english]{babel}
\usepackage{graphicx,color}
\usepackage{epsfig}
\usepackage{bm}
\usepackage{longtable}
\usepackage{verbatim} 
\usepackage{float}
\usepackage{longtable}
\usepackage{multirow}
\usepackage[utf8]{inputenc}

\newcommand{\phidag}{\phi^{\dagger}}
\newcommand{\vdag[1]}{V^{\dagger}_{#1}}
\newcommand{\Vdag[1]}{V^{#1\dagger}}

\newcommand{\ld[1]}{\lambda_{#1}}

\newcommand{\mathsym}[1]{{}}

\input{tcilatex}
\begin{document}

\title{Generating lepton masses and mixings with a heavy vector doublet.}
\author{A. E. C\'arcamo Hern\'andez}
\email{antonio.carcamo@usm.cl}
\author{Jonatan Vignatti}
\email{jonatan.vignatti@sansano.usm.cl}
\author{Alfonso Zerwekh}
\email{alfonso.zerwekh@usm.cl}
\affiliation{Universidad T\'{e}cnica Federico Santa Mar\'{\i}a\\
and Centro Cient\'{\i}fico-Tecnol\'{o}gico de Valpara\'{\i}so\\
Casilla 110-V, Valpara\'{\i}so, Chile,}
\date{\today }

\begin{abstract}
We construct two viable extensions of the SM with a heavy vector in the
fundamental $SU\left( 2\right) _{L}$ representation and nine SM singlet
scalar fields, consistent with the current SM fermion mass spectrum and
fermionic mixing parameters. The small masses for the active neutrinos are
generated from radiative seesaw mechanism at one loop level mediated by the
neutral components of the heavy vector as well as by the left handed
Majorana neutrinos. The proposed models predicts rates for charged lepton flavor violating processes within the reach of the forthcoming experiments.
\end{abstract}

\maketitle

\section{Introduction}

In spite of its remarkable agreement with the experimental data, the
Standard Model (SM) leaves unexplained several issues such as, for example,
the number of fermion families, the observed hierarchical fermion mass
spectrum and mixing angles, etc. Whereas the CKM quark mixing matrix is
nearly diagonal thus implying small quark mixing angles, the PMNS leptonic
mixing matrix significantly deviates from the identity matrix thus giving
rise to sizeable leptonic mixing angles. The experimental data indicates
that two of the leptonic mixing angles are large whereas one mixing angle is
Cabibbo sized, which contrast with the small values of the CKM paramters. In
addition, the pattern of fermion masses in the SM is extended over a range
of about thirteen orders of magnitude ranging from the light active neutrino mass scale up
to the top quark mass. This is so called SM flavor puzzle which is left unaddressed in the SM.

That SM ``flavor puzzle'' motivates to build models with additional scalars
and fermions in their particle spectrum and with an extended gauge group,
supplemented by discrete flavour symmetries, which are usually spontaneously
broken, in order to generate the observed pattern of SM fermion masses and
mixing angles. Recent reviews of discrete flavor groups can be found in
Refs. \cite%
{Ishimori:2010au,Altarelli:2010gt,King:2013eh,King:2014nza,King:2017guk}.
Several discrete groups such as $S_{3}$ \cite%
{Gerard:1982mm,Kubo:2003pd,Kobayashi:2003fh,Chen:2004rr,Mondragon:2007af,Mondragon:2008gm,Bhattacharyya:2010hp, Dong:2011vb,Dias:2012bh,Meloni:2012ci,Canales:2012dr,Canales:2013cga,Ma:2013zca,Kajiyama:2013sza,Hernandez:2013hea, Ma:2014qra,Hernandez:2014vta,Hernandez:2014lpa,Gupta:2014nba,Hernandez:2015dga,Hernandez:2015zeh,Hernandez:2016rbi,Hernandez:2015hrt,CarcamoHernandez:2016pdu,Arbelaez:2016mhg,Gomez-Izquierdo:2017rxi,Cruz:2017add}, $A_{4}$ \cite%
{Ma:2001dn,He:2006dk,Feruglio:2008ht,Feruglio:2009hu,Chen:2009um,Varzielas:2010mp,Altarelli:2012bn,Ahn:2012tv,Memenga:2013vc,Felipe:2013vwa,Varzielas:2012ai, Ishimori:2012fg,King:2013hj,Hernandez:2013dta,Babu:2002dz,Altarelli:2005yx,Gupta:2011ct,Morisi:2013eca, Altarelli:2005yp,Kadosh:2010rm,Kadosh:2013nra,delAguila:2010vg,Campos:2014lla,Vien:2014pta,Joshipura:2015dsa,Hernandez:2015tna,Karmakar:2016cvb,Chattopadhyay:2017zvs,CarcamoHernandez:2017kra,Ma:2017moj,CentellesChulia:2017koy,Bjorkeroth:2017tsz,Srivastava:2017sno,Borah:2017dmk,Belyaev:2018vkl,CarcamoHernandez:2018aon,Srivastava:2018ser,delaVega:2018cnx,Borah:2018nvu,Pramanick:2019qpg,CarcamoHernandez:2019pmy,CarcamoHernandez:2019kjy}%
, $S_{4}$ \cite%
{Bazzocchi:2009da,Bazzocchi:2009da,Altarelli:2009gn,Toorop:2010yh,Patel:2010hr,Morisi:2011pm,Mohapatra:2012tb,BhupalDev:2012nm,Varzielas:2012pa,Ding:2013hpa,Ishimori:2010fs,Ding:2013eca,Hagedorn:2011un,Campos:2014zaa,Dong:2010zu,VanVien:2015xha,deAnda:2017yeb,deAnda:2018oik,CarcamoHernandez:2019eme,CarcamoHernandez:2019iwh}, $D_{4}$ \cite%
{Frampton:1994rk,Grimus:2003kq,Grimus:2004rj,Frigerio:2004jg,Adulpravitchai:2008yp,Ishimori:2008gp,Hagedorn:2010mq,Meloni:2011cc,Vien:2013zra}%
, $Q_{6}$ \cite%
{Babu:2004tn,Kajiyama:2005rk,Kajiyama:2007pr,Kifune:2007fj,Babu:2009nn,
Kawashima:2009jv,Kaburaki:2010xc,Babu:2011mv,Araki:2011zg,
Gomez-Izquierdo:2013uaa,Gomez-Izquierdo:2017med}, $T_{7}$ \cite{Luhn:2007sy,Hagedorn:2008bc,Cao:2010mp,Luhn:2012bc,Kajiyama:2013lja,Bonilla:2014xla,Vien:2014gza, Vien:2015koa,Hernandez:2015cra,Arbelaez:2015toa}, $T_{13}$ \cite{Ding:2011qt,Hartmann:2011dn,Hartmann:2011pq,Kajiyama:2010sb}, $T^{\prime }$ \cite{Aranda:2000tm,Feruglio:2007uu,Sen:2007vx,Aranda:2007dp,Chen:2007afa,Eby:2008uc,Frampton:2008bz,Frampton:2008ep,Eby:2009ii,Frampton:2009fw,Merlo:2011hw,Eby:2011ph,Eby:2011qa,Chen:2011tj,Meroni:2012ty,Frampton:2013lva,Chen:2013wba,Girardi:2013sza,Carone:2016xsi,Vien:2018otl,Carone:2019lfc,CarcamoHernandez:2019vih}, $\Delta (27)$ \cite%
{Branco:1983tn,deMedeirosVarzielas:2006fc,Ma:2007wu,Varzielas:2012nn,Bhattacharyya:2012pi,Ferreira:2012ri,Ma:2013xqa,Nishi:2013jqa,Varzielas:2013sla,Aranda:2013gga,Harrison:2014jqa,Ma:2014eka,Abbas:2014ewa,Abbas:2015zna,Varzielas:2015aua,Bjorkeroth:2015uou,Chen:2015jta,Vien:2016tmh,Hernandez:2016eod,CarcamoHernandez:2017owh,deMedeirosVarzielas:2017sdv,Bernal:2017xat,CarcamoHernandez:2018iel,deMedeirosVarzielas:2018vab,CarcamoHernandez:2018hst,CarcamoHernandez:2018djj,Bjorkeroth:2019csz}, $\Delta(96)$ \cite{King:2012in,King:2013vna,Ding:2014ssa}, $\Delta(6N^2)$ \cite{Ishimori:2014jwa,King:2014rwa,Ishimori:2014nxa} and $A_{5}$ \cite%
{Everett:2008et,Feruglio:2011qq,Cooper:2012bd,Varzielas:2013hga,Gehrlein:2014wda,Gehrlein:2015dxa,DiIura:2015kfa,Ballett:2015wia,Gehrlein:2015dza,Turner:2015uta,Li:2015jxa,Ding:2017hdv}
have been implemented in extensions of the SM, to provide a nice description
of the observed pattern of fermion masses and mixing angles.

On the other hand, given the current lack of experimental evidence in favor
of the traditional big paradigms of Physics beyond the Standard Model, it
seems prudent to explore more exotic paths. In recent years, for instance,
some groups have pay attention to spin-1 fields transforming in the
fundamental representation of $SU(2)_L$\cite{Chizhov:2009fc}. This kind of
field may naturally appear, for instance, in models such as: Higgs-Gauge
Unification\cite{Maru:2018ocf} and Composite Higgs\cite{Agashe:2009bb,Franzosi:2016aoo}. 
In a previous paper, our group has studied the phenomenology of a spin-1 doublet which has the
same hypercharge of the Higgs doublet \cite{Saez:2018off}. In the context of Composite Higgs, this kind of field may be interpreted as a spin-1 excitation of the Higgs field in analogy
to the rho-mesons which can be seen as the spin-1 excitation of the pions. Such a spin-1 doublet cannot be coupled to standard fermions except by the introduction of exotic fields. The simplest alternative is the introduction of a left-handed fermion which is singlet under the SM group. Of course, such an exotic fermion will behave like a sterile neutrino. In this case, the spin-1 doublet will couple the standard leptons to new exotic neutrino. The
aim of this work is to show that the introduction of such new fields (the spin-1 doublet and the new sterile left-handed neutrino) can have an impact on
neutrino Physics by providing a new mechanism for the mass generation of the
light active neutrinos.

In this paper, we extend the Standard Model by introducing a new vector
doublet field. We treat it as matter field and the resulting model is
thought as an effective theory. In other words, we will not discuss the
details of the dynamics that generates this vector doublet field and its
mass which is beyond the scope of this article. The consistency of such a
construction with perturbative unitarity and collider phenomenology have
been studied elsewhere \cite{Saez:2018off}.

Additionally, we supplement SM gauge symmetry by including a $S_3$ discrete group as well as several cyclic symmetries
%$S_{3}\times
%Z_{2}\times Z_{6}\times Z_{8}\times Z_{12}$ discrete group 
and we extend the particle
content by adding several gauge singlet %is extended to include nine SM 
scalars, two left handed
neutrinos $N_{nL}(n=1,2)$, singlets under the SM gauge group and a $SU\left(
2\right) _{L}$ doublet of heavy vectors. We introduce several gauge singlet scalar fields to provide an explanation for the observed hierarchies in the SM fermion mass spectrum and the fermion mixing parameters while keeping all the Yukawa couplings of order unity. We consider two models: The first one is valid only for the inverted neutrino mass hierarchy whereas the second one allows to successfully account for lepton masses and mixing in the scenario of normal neutrino mass ordering. These models are consistent with the SM
fermion masses and mixings. The effective neutrino mass matrix arises
through radiative seesaw.

%In the model under consideration the light active neutrino masses arise from a one loop level radiative seesaw mechanism mediated by the neutral components of the heavy vector and the left handed Majorana neutrinos.

The paper is organized as follows. In section \ref{model} we explain the first model. In Sec. \ref{quarksector} we focus on the discussion of quark masses
and mixing and give our corresponding results. In Sec. \ref{leptonsector} we
discuss the implications of our model on lepton masses and mixings. In section \ref{alter} we propose an alternative model in order to account for lepton masses and mixings for the case of normal neutrino mass hierarchy. Section \ref{LFV} provides a discussion of the charged lepton flavor violation constraints on the model parameter space. We conclude in section \ref{conclusions}. Appendix \ref{S3} provides a concise description of the $S_{3}$ discrete group.

\section{The model}

\label{model}

We propose an extension of the Standard Model (SM) where the SM gauge
symmetry is supplemented by the $S_{3}\times Z_{2}\times Z_{6}\times
Z_{8}\times Z_{12}$ discrete group and the particle content is extended to
include the SM scalars singlets $\varphi ,\chi ,\xi ,\eta ,\sigma ,\rho
_{k}(k=1,2)$, two left handed neutrinos $N_{nL}(n=1,2)$, singlets under the
SM gauge group and a $SU\left( 2\right) _{L}$ doublet of heavy vectors. The
full symmetry $\mathcal{G}$ of our model experiences the following two step
spontaneous breaking:%
\begin{eqnarray}
&&\mathcal{G}=SU\left( 3\right) _{C}\times SU\left( 2\right) _{L}\times
U\left( 1\right) _{Y}\times S_{3}\times Z_{2}\times Z_{6}\times Z_{8}\times
Z_{12}  \label{Group} \\
&&\hspace{35mm}\Downarrow \Lambda _{int}  \notag \\[0.12in]
&&\hspace{15mm}SU(3)_{C}\times SU\left( 2\right) _{L}\times U\left( 1\right)
_{Y}  \notag \\[0.12in]
&&\hspace{35mm}\Downarrow v  \notag \\[0.12in]
&&\hspace{15mm}SU(3)_{C}\times U\left( 1\right) _{Q}  \notag
\end{eqnarray}%
where the symmetry breaking scales satisfy the hierarchy $\Lambda_{int}>v$
and $v=246$ GeV is the electroweak symmetry breaking scale.

The vector doublet is described by the following Lagrangian:
\begin{eqnarray} \label{eq:fl}
{\cal L} &=& -\frac{1}{2}\left(D_{\mu} V_{\nu} - D_{\nu} V_{\mu} \right)^{\dagger} \left(D^{\mu} V^{\nu} - D^{\nu} V^{\mu} \right) +  M_V^2 \vdag[\mu] V^{\mu} \nonumber \\
& + &\ld[2](\phidag \phi) (\vdag[\mu]V^{\mu}) + \ld[3](\phidag V_{\mu})(\Vdag[\mu]\phi)  + \ld[4](\phidag V_{\mu})(\phidag V^{\mu}) \nonumber \\ 
\nonumber &+&  \alpha_1 \phidag D_{\mu} V^{\mu} + \alpha_2 (\vdag[\mu] V^{\mu}) (\vdag[\nu] V^{\nu}) + \alpha_3 (\vdag[\mu] V^{\nu}) (\vdag[\nu] V^{\mu}) \nonumber \\
&+& ig\kappa_1V_\mu^\dagger W^{\mu\nu} V_\nu + i\frac{g'}{2}\kappa_2 V_\mu^\dagger B^{\mu\nu} V_\nu  + h.c.
\end{eqnarray}
where $\phi$ is the standard Higgs doublet and $V_{\mu}$ is the new vector doublet.

Notice that the dimension-3 term $\alpha_1 \phidag D_{\mu} V^{\mu}$ can be safely dropped out (or, equivalently, we can set $\alpha_1=0$) because in that case an accidental $Z_2$ symmetry opens up. In this way, the mixing of the new vectors and the gauge boson is forbidden at tree level. The consequences of this Lagrangian (with $\alpha_1=0$) in absence of any other New Physics was already studied in detail in Ref. \cite{Saez:2018off}. In this work, we will explore the implications of these new vectors in the generation of masses for light active neutrinos. 

In our model the fermion sector is augmented by adding two heavy
gauge singlet left handed Majorana neutrinos. The transformation properties of the 
quark, lepton and scalar fields under the $S_{3}\times Z_{2}\times
Z_{6}\times Z_{8}\times Z_{12}$ discrete group are indicated in Tables \ref%
{quarks}, \ref{leptons} and \ref{scalars}, respectively.

\begin{table}[tbp]
\centering 
\begin{tabular}{|c|c|c|c|c|c|c|c|c|c|}
\hline
& $q_{1L}$ & $q_{2L}$ & $q_{3L}$ & $u_{1R}$ & $u_{2R}$ & $u_{3R}$ & $d_{1R}$
& $d_{2R}$ & $d_{3R}$ \\ \hline
$S_3$ & $\mathbf{1}$ & $\mathbf{1}$ & $\mathbf{1}$ & $\mathbf{1}$ & $\mathbf{%
1}$ & $\mathbf{1}$ & $\mathbf{1^{\prime}}$ & $\mathbf{1^{\prime}}$ & $%
\mathbf{1^{\prime}}$ \\ \hline
$Z_2$ & $0$ & $0$ & $0$ & $0$ & $0$ & $0$ & $0$ & $0$ & $0$ \\ \hline
$Z_6$ & $0$ & $0$ & $0$ & $0$ & $0$ & $0$ & $3$ & $3$ & $3$ \\ \hline
$Z_8$ & $-2$ & $-1$ & $0$ & $2$ & $1$ & $0$ & $2$ & $1$ & $0$ \\ \hline
$Z_{12}$ & $0$ & $0$ & $0$ & $6$ & $6$ & $0$ & $0$ & $0$ & $0$ \\ \hline
\end{tabular}%
\caption{Transformation properties of the quark fields under the $%
S_{3}\times Z_{2}\times Z_{6}\times Z_{8}\times Z_{12}$ discrete group.}
\label{quarks}
\end{table}
\begin{table}[tbp]
\centering%
\begin{tabular}{|c|c|c|c|c|c|c|c|}
\hline
& $l_{1L}$ & $l_{L}$ & $l_{1R}$ & $l_{2R}$ & $l_{3R}$ & $N_{1L}$ & $N_{2L}$
\\ \hline
$S_{3}$ & $\mathbf{1^{\prime }}$ & $\mathbf{2}$ & $\mathbf{1}$ & $\mathbf{1}$
& $\mathbf{1^{\prime }}$ & $\mathbf{1}$ & $\mathbf{1}$ \\ \hline
$Z_{2}$ & $0$ & $0$ & $0$ & $0$ & $0$ & $1$ & $1$ \\ \hline
$Z_{6}$ & $-2$ & $-2$ & $-2$ & $-2$ & $-2$ & $3$ & $3$ \\ \hline
$Z_{8}$ & $-3$ & $-1$ & $-1$ & $3$ & $-1$ & $4$ & $4$ \\ \hline
$Z_{12}$ & $0$ & $0$ & $0$ & $0$ & $6$ & $6$ & $6$ \\ \hline
\end{tabular}%
\caption{Transformation properties of the leptonic fields under the $%
S_{3}\times Z_{2}\times Z_{6}\times Z_{8}\times Z_{12}$ discrete group.}
\label{leptons}
\end{table}
\begin{table}[tbp]
\centering 
\begin{tabular}{|c|c|c|c|c|c|c|c|c|}
\hline
& $\phi$ & $\varphi$ & $\chi$ & $\xi$ & $\eta$ & $\sigma$ & $\rho_1$ & $%
\rho_2$ \\ \hline
$S_3$ & $\mathbf{1}$ & $\mathbf{1}$ & $\mathbf{2}$ & $\mathbf{2}$ & $\mathbf{%
1}$ & $\mathbf{1}$ & $\mathbf{1}$ & $\mathbf{1^{\prime}}$ \\ \hline
$Z_2$ & $0$ & $0$ & $0$ & $0$ & $0$ & $0$ & $0$ & $0$ \\ \hline
$Z_6$ & $0$ & $-2$ & $0$ & $-3$ & $0$ & $-3$ & $0$ & $0$ \\ \hline
$Z_8$ & $0$ & $-1$ & $0$ & $0$ & $-1$ & $-2$ & $0$ & $0$ \\ \hline
$Z_{12}$ & $0$ & $0$ & $0$ & $0$ & $0$ & $0$ & $-3$ & $-2$ \\ \hline
\end{tabular}%
\caption{Transformation properties of the scalar fields under the $%
S_{3}\times Z_{2}\times Z_{6}\times Z_{8}\times Z_{12}$ discrete groups.}
\label{scalars}
\end{table}
Here the dimensions of the $S_{3}$ irreducible representations are specified
by the numbers in boldface and the different $Z_{2}\times Z_{6}\times
Z_{8}\times Z_{12}$ charges are written in additive notation.

The $S_{3}\times Z_{2}\times Z_{6}\times Z_{8}\times Z_{12}$ assignment of
the heavy vector in the fundamental $SU\left( 2\right) _{L}$ representation
is: 
\begin{equation}
V_{\mu }\sim \left( \mathbf{1},1,0,0,0\right) .
\end{equation}%
%
%
%
%
%
%
%
%
% The $S_{3}\times Z_{2}\times Z_{6}\times Z_{8}\times Z_{12}$ assignments of
% the scalar fields are: 
From the point of view of the gauge symmetry, $V_{\mu}$ is assumed to
transform as a matter field in the fundamental representation of $SU(2)_L$, 
\textit{i.e.} $V_{\mu}\rightarrow UV_{\mu}$ with $U\in SU(2)_L$.
Additionally, we assume that the hypercharge of the new vector double is the
same as the SM Higgs doublet. As a consequence, as we will see below, our
model results to be similar to the well known Two Higgs Doublet Model with
the second scalar doublet replaced by a vector one.

With the particle spectrum previously specified, we get the following quark
and lepton Yukawa terms as well as the interaction terms of the heavy vector
with the heavy left handed Majorana neutrinos: 
\begin{eqnarray}
\tciLaplace _{Y}^{\left( q\right) } &=&y_{33}^{\left( u\right) }\overline{q}%
_{3L}\widetilde{\phi }u_{3R}+y_{23}^{\left( u\right) }\overline{q}_{2L}%
\widetilde{\phi }u_{3R}\frac{\eta }{\Lambda }+y_{13}^{\left( u\right) }%
\overline{q}_{1L}\widetilde{\phi }u_{3R}\frac{\eta ^{2}}{\Lambda ^{2}} 
\notag \\
&&+y_{32}^{\left( u\right) }\overline{q}_{3L}\widetilde{\phi }u_{2R}\frac{%
\eta \rho _{1}^{2}}{\Lambda ^{3}}+y_{22}^{\left( u\right) }\overline{q}_{2L}%
\widetilde{\phi }u_{2R}\frac{\eta ^{2}\rho _{1}^{2}}{\Lambda ^{4}}%
+y_{12}^{\left( u\right) }\overline{q}_{1L}\widetilde{\phi }u_{2R}\frac{\eta
^{3}\rho _{1}^{2}}{\Lambda ^{5}}  \notag \\
&&+y_{11}^{\left( u\right) }\overline{q}_{1L}\widetilde{\phi }u_{1R}\frac{%
\eta ^{2}\rho _{1}^{2}}{\Lambda ^{4}}+y_{21}^{\left( u\right) }\overline{q}%
_{2L}\widetilde{\phi }u_{1R}\frac{\eta ^{3}\rho _{1}^{2}}{\Lambda ^{5}}%
+y_{11}^{\left( u\right) }\overline{q}_{1L}\widetilde{\phi }u_{1R}\frac{\eta
^{4}\rho _{1}^{2}}{\Lambda ^{6}}  \notag \\
&&+\sum_{j=1}^{3}\sum_{k=1}^{3}y_{jk}^{\left( d\right) }\overline{q}%
_{jL}\phi d_{kR}\frac{\eta ^{6-j-k}\left( \xi \xi \xi \right) _{\mathbf{%
\mathbf{1}^{\prime }}}}{\Lambda ^{9-j-k}}+h.c,  \label{Lyq}
\end{eqnarray}%
\begin{eqnarray}
\tciLaplace _{Y}^{\left( l\right) } &=&y_{11}^{\left( l\right) }\overline{l}%
_{1L}\phi l_{1R}\frac{\eta ^{2}\rho _{2}^{3}\varphi ^{4}}{\Lambda ^{9}}%
+y_{22}^{\left( l\right) }\left( \overline{l}_{L}\phi \chi \right) _{\mathbf{%
\mathbf{1}}}l_{2R}\frac{\eta ^{4}}{\Lambda ^{5}}+y_{13}^{\left( l\right) }%
\overline{l}_{1L}\phi l_{3R}\frac{\eta ^{2}\rho _{1}^{2}}{\Lambda ^{4}}%
+y_{23}^{\left( l\right) }\left( \overline{l}_{L}\phi \chi \right) _{\mathbf{%
\mathbf{1}}}l_{3R}\frac{\rho _{2}^{3}}{\Lambda ^{4}}  \notag \\
&&+y_{33}^{\left( l\right) }\left( \overline{l}_{L}\phi \chi \right) _{%
\mathbf{\mathbf{1}^{\prime }}}l_{3R}\frac{\rho _{1}^{2}}{\Lambda ^{3}}+\frac{%
1}{2}\sum_{n=1}^{2}m_{N_{n}}\overline{N}_{nL}N_{nL}^{C}+h.c,  \label{Lyl}
\end{eqnarray}%
\begin{equation}
-\mathcal{L}_{VlN}=\sum_{n=1}^{2}y_{1n}^{\left( V\right) }\overline{l}%
_{1L}\gamma ^{\mu }V_{\mu }N_{nL}\frac{\sigma \eta ^{4}\rho _{1}^{2}\varphi 
}{\Lambda ^{8}}+\sum_{n=1}^{2}y_{2n}^{\left( V\right) }\overline{l}%
_{L}\gamma ^{\mu }V_{\mu }N_{nL}\frac{\xi \eta ^{4}\rho _{1}^{2}\varphi }{%
\Lambda ^{8}}  \label{LVI}
\end{equation}%
where $y_{ij}^{\left( u,d\right) }$ ($i,j=1,2,3$), $y_{11}^{\left( l\right)
} $, $y_{22}^{\left( l\right) }$, $y_{13}^{\left( l\right) }$, $%
y_{23}^{\left( l\right) }$, $y_{33}^{\left( l\right) }$, $y_{1n}^{\left(
V\right) }$ and $y_{2n}^{\left( V\right) }$ are dimensionless quantities.
Notice that, given the quantum number assignments on the discrete group
factors and the transformation properties of the fields under the gauge
symmetry, the coupling written in eq. (\ref{LVI}) are the only ones allowed
involving the new vector doublet and fermions (excepting for higher
dimensional operators).

From the interactions terms of the heavy vector with the heavy left handed
Majorana neutrinos we find: 
\begin{equation}
-\mathcal{L}_{VlN}\supset \sum_{n=1}^{2}y_{1n}^{\left( V\right) }\overline{l}%
_{1L}\gamma ^{\mu }V_{\mu }N_{nL}\frac{v_{\sigma }v_{\eta }^{4}v_{\rho
_{1}}^{2}v_{\varphi }}{\Lambda ^{8}}+\sum_{n=1}^{2}y_{2n}^{\left( V\right)
}\left( \overline{l}_{2L}+\sqrt{2}\overline{l}_{3L}\right) \gamma ^{\mu
}V_{\mu }N_{nL}\frac{v_{\xi }v_{\eta }^{4}v_{\rho _{1}}^{2}v_{\varphi }}{%
\Lambda ^{8}}.  \label{LyV}
\end{equation}%
Besides that, as the hierarchy among charged fermion masses and quark mixing
angles mass emerges from the breaking of the $S_{3}\times Z_{6}\times
Z_{8}\times Z_{12}$ discrete group, we set the VEVs of the SM singlet scalar
fields with respect to the Wolfenstein parameter $\lambda =0.225$, and the
model cutoff $\Lambda $, as follows: 
\begin{equation}
v_{\eta }\sim v_{\rho _{1}}\sim v_{\rho _{2}}\sim v_{\varphi }\sim v_{\chi
}\sim v_{\xi }=v_{\sigma }=\Lambda _{int}=\lambda \Lambda .  \label{VEV}
\end{equation}%
The role of the different discrete group factors of the model is explained
in the following. The $S_{3}$, $Z_{6}$, $Z_{8}$ and $Z_{12}$ discrete groups
allow to reduce the number of model parameters and set the SM charged lepton
mass hierarchy, which is crucial to get viable textures for the lepton
sector consistent with the current pattern of lepton masses and mixings, as
we will show in Section \ref{leptonsector}. We use the $S_{3}$ discrete
group because since it is the smallest non-Abelian group that has been
considerably studied in the literature. The $S_{3}$, $Z_{6}$, $Z_{8}$ and $%
Z_{12}$ symmetries determine the allowed entries of the charged lepton mass
matrix. In addition, the $Z_{6}$ symmetry separates the $S_{3}$ scalar
doublet $\chi $ participating in the charged lepton Yukawa interactions from
the $S_{3}$ doublet $\xi $ that appear in the neutrino Yukawa terms.
Furthermore, the $Z_{8}$ and $Z_{12}$ symmetries are crucial for explaining
the tau and muon lepton masses and for providing the Cabbibo sized value for
the reactor mixing angle $\theta _{13}$ as well as the Cabbibo sized
corrections to the atmospheric mixing angle $\theta _{23}$, without tuning
the charged lepton Yukawa couplings. The smallness of the electron mass is
explained by the $S_{3}$, $Z_{6}$, $Z_{8}$ and $Z_{12}$\ discrete
symmetries. We use the $Z_{12}$ discrete symmetry since it is the smallest
cyclic symmetry that glue $\rho _{1}^{2}$ and $\rho _{2}^{3}$ with $l_{3R}$,
considering $l_{3R}$ charged under this symmetry. The $Z_{2}$ symmetry,
under which only the Majorana neutrinos and the heavy vector are charged, is
introduced in order to avoid a tree level type I seesaw mechanism for the
generation of the light active neutrino masses.

In order to get predictive and viable fermion sector, we assume that the $%
S_{3}$ doublet SM singlet scalars $\chi$ and $\xi$ have the following VEV
configurations: 
\begin{equation}
\left\langle \chi \right\rangle =v_{\chi }\left( 1,0\right) ,\hspace{1cm}%
\hspace{1cm}\left\langle \xi \right\rangle =v_{\xi }\left( 1,\sqrt{2}\right).
\end{equation}%
The VEV configurations given above correspond to natural solutions of the
minimization equations of the scalar potential for the whole region of
parameter space as shown in Ref. \cite{Hernandez:2015dga}.

\section{Quark masses and mixings}

\label{quarksector}

From the quark Yukawa terms of Eq. (\ref{Lyq}), we get that mass matrices
for the SM quarks are: 
\begin{equation}
M_{U}=\left( 
\begin{array}{ccc}
a_{11}^{\left( u\right) }\lambda ^{6} & a_{12}^{\left( u\right) }\lambda ^{5}
& a_{13}^{\left( u\right) }\lambda ^{2} \\ 
a_{21}^{\left( u\right) }\lambda ^{5} & a_{22}^{\left( u\right) }\lambda ^{4}
& a_{23}^{\left( u\right) }\lambda \\ 
a_{31}^{\left( u\right) }\lambda ^{4} & a_{32}^{\left( u\right) }\lambda ^{3}
& a_{33}^{\left( u\right) }%
\end{array}%
\right) \frac{v}{\sqrt{2}},\hspace{1cm}\hspace{1cm}M_{D}=\left( 
\begin{array}{ccc}
a_{11}^{\left( d\right) }\lambda ^{7} & a_{12}^{\left( d\right) }\lambda ^{6}
& a_{13}^{\left( d\right) }\lambda ^{5} \\ 
a_{21}^{\left( d\right) }\lambda ^{6} & a_{22}^{\left( d\right) }\lambda ^{5}
& a_{23}^{\left( d\right) }\lambda ^{4} \\ 
a_{31}^{\left( d\right) }\lambda ^{5} & a_{32}^{\left( d\right) }\lambda ^{4}
& a_{33}^{\left( d\right) }\lambda ^{3}%
\end{array}%
\right) \frac{v}{\sqrt{2}},  \label{Mq}
\end{equation}%
with $\lambda =0.225$ being one of the Wolfenstein parameters, $v=246$ GeV
the scale of breaking of the electroweak gauge symmetry and $a_{ij}^{\left(
u,d\right) }$ ($i,j=1,2,3$) are $\mathcal{O}(1)$ parameters. Since the
charged fermion mass and quark mixing pattern is caused by the spontaneous
breaking of the $S_{3}\times Z_{6}\times Z_{8}\times Z_{12}$ discrete group
and in order to simplify the analysis, we adopt the following scenario: 
\begin{eqnarray}
a_{12}^{\left( u\right) } &=&a_{21}^{\left( u\right) },\hspace{0.7cm}%
a_{31}^{\left( u\right) }=a_{13}^{\left( u\right) },\hspace{0.7cm}%
a_{32}^{\left( u\right) }=a_{23}^{\left( u\right) },  \notag \\
a_{12}^{\left( d\right) } &=&\left\vert a_{12}^{\left( d\right) }\right\vert
e^{-i\tau _{1}},\hspace{0.75cm}a_{21}^{\left( d\right) }=\left\vert
a_{12}^{\left( d\right) }\right\vert e^{i\tau _{1}}, \\
a_{13}^{\left( d\right) } &=&\left\vert a_{13}^{\left( d\right) }\right\vert
e^{-i\tau _{2}},\hspace{0.75cm}a_{31}^{\left( d\right) }=\left\vert
a_{13}^{\left( d\right) }\right\vert e^{i\tau _{2}},\hspace{0.75cm}%
a_{23}^{\left( d\right) }=a_{32}^{\left( d\right) }.  \notag
\end{eqnarray}%
Furthermore we set $a_{33}^{\left( u\right) }=1$, which is suggested by
naturalness arguments. Notice that we have chosen the unusual scaling of quark mass textures in Eq. (\ref{Mq}) in order to avoid adding more symmetries and scalar fields in our model, that will be required to get several zero entries in the SM quark mass matrices, which are needed to get a more natural scaling in terms of powers of the Wolfenstein parameter.

For the quark mass matrices given above and considering the benchmark
scenario previously described, we look for the eigenvalue problem solutions
reproducing the experimental values of the quark masses \cite%
{Bora:2012tx,Xing:2007fb}, quark mixing parameters and CP violating phase 
\cite{Olive:2016xmw}, under the condition that the effective parameters $%
a_{ij}^{\left( u,d\right) }$ ($i,j=1,2,3$) be most close to $\mathcal{O}(1)$%
. Applying the standard procedure we find the following solution: 
\begin{eqnarray}
a_{11}^{\left( u\right) } &\simeq &0.58,\hspace{1cm}a_{22}^{\left( u\right)
}\simeq 2.19,\hspace{1cm}a_{12}^{\left( u\right) }\simeq 0.67,  \notag \\
a_{13}^{\left( u\right) } &\simeq &0.80,\hspace{1cm}a_{23}^{\left( u\right)
}\simeq 0.83,\hspace{1cm}a_{11}^{\left( d\right) }\simeq 1.96,  \notag \\
a_{12}^{\left( d\right) } &\simeq &0.53,\hspace{1cm}a_{13}^{\left( d\right)
}\simeq 1.07,\hspace{1cm}a_{22}^{\left( d\right) }\simeq 1.93, \\
a_{23}^{\left( d\right) } &\simeq &1.36,\hspace{1cm}a_{33}^{\left( d\right)
}\simeq 1.35,\hspace{1.2cm}\tau _{1}\simeq 9.56^{\circ },\hspace{0.5cm}\tau
_{2}\simeq 4.64^{\circ }.  \notag
\end{eqnarray}%
\begin{table}[tbh]
\begin{center}
\begin{tabular}{c|l|l}
\hline\hline
Observable & Model value & Experimental value \\ \hline
$m_{u}(MeV)$ & \quad $1.44$ & \quad $1.45_{-0.45}^{+0.56}$ \\ \hline
$m_{c}(MeV)$ & \quad $656$ & \quad $635\pm 86$ \\ \hline
$m_{t}(GeV)$ & \quad $177.1$ & \quad $172.1\pm 0.6\pm 0.9$ \\ \hline
$m_{d}(MeV)$ & \quad $2.9$ & \quad $2.9_{-0.4}^{+0.5}$ \\ \hline
$m_{s}(MeV)$ & \quad $57.7$ & \quad $57.7_{-15.7}^{+16.8}$ \\ \hline
$m_{b}(GeV)$ & \quad $2.82$ & \quad $2.82_{-0.04}^{+0.09}$ \\ \hline
$\sin \theta _{12}$ & \quad $0.225$ & \quad $0.225$ \\ \hline
$\sin \theta _{23}$ & \quad $0.0412$ & \quad $0.0412$ \\ \hline
$\sin \theta _{13}$ & \quad $0.00351$ & \quad $0.00351$ \\ \hline
$\delta $ & \quad $64^{\circ }$ & \quad $68^{\circ }$ \\ \hline\hline
\end{tabular}%
\end{center}
\caption{Model and experimental values of the quark masses and CKM
parameters.}
\label{Tabquarks}
\end{table}
We use the $M_{Z}$-scale experimental values of the quark masses given by Ref. \cite{Bora:2012tx}
(which are similar to those in \cite{Xing:2007fb}). The experimental values
of the CKM parameters are taken from Ref. \cite{Olive:2016xmw}. As shown in Table \ref{Tabquarks}, the quark mass spectrum, quark mixing parameters and CP violating phase obtained in our model are in very good agreement with the experimental data.
%As shown in Table \ref{Tabquarks}, the quark mass spectrum, \cite{Bora:2012tx,Xing:2007fb}, quark mixing parameters and CP violating phase \cite{Olive:2016xmw} are in very good agreement with the experimental data.

\section{Lepton masses and mixings}

\label{leptonsector} 
\begin{figure}[tbp]
\includegraphics[scale=0.5, angle=0]{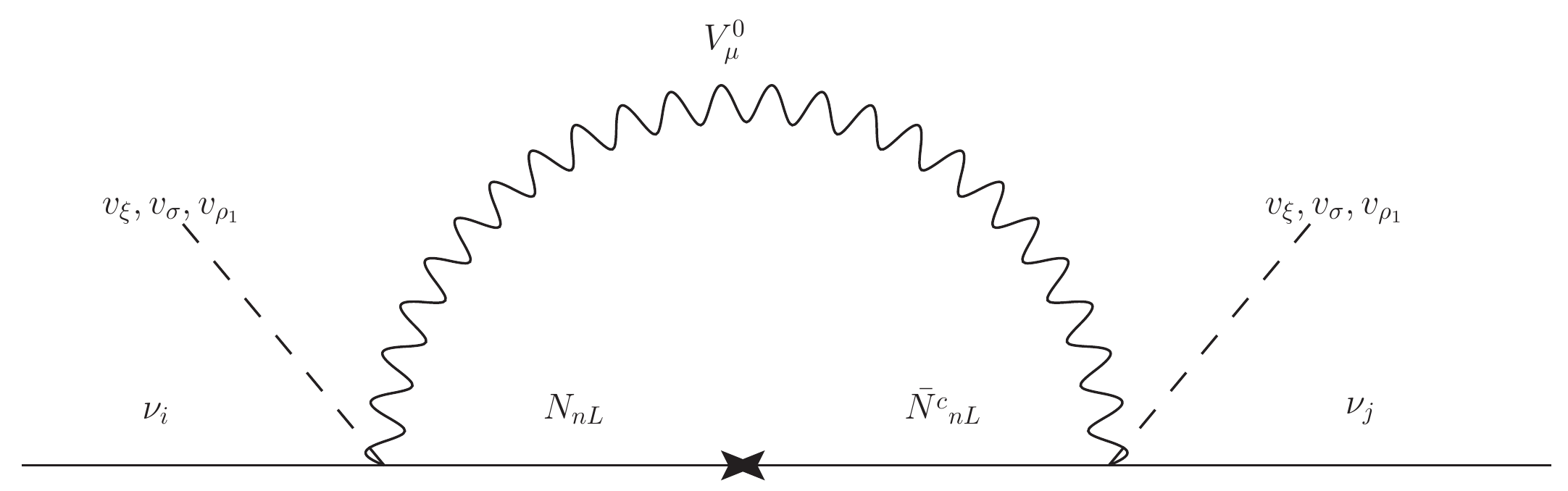}
\caption{Loop diagram illustrating the generation of the light active
neutrino mass. Here $n=1,2$ and $j=1,2,3$.}
\label{fig:diagram}
\end{figure}

From the interactions terms of the heavy vector with
the heavy left handed Majorana neutrinos of Eq. (\ref{LyV}), it follows that
light active neutrino masses are generated from a one loop level radiative
seesaw mechanism mediated by the left handed Majorana neutrinos and by the
real and imaginary parts of the neutral components of the heavy vector in
the fundamental $SU\left( 2\right) _{L}$ representation, as illustrated in
Fig. \ref{fig:diagram}. 
Thus, the elements of the light active neutrino mass matrix take the form: 
\begin{eqnarray}
\left( M_{\nu }\right) _{11} &=&\sum_{n=1}^{2}\left( y_{1n}^{\left( V\right)
}\right) ^{2}f\left( m_{\func{Re}V^{0}},m_{\func{Im}V^{0}},m_{N_{n}}\right)
\left( \frac{v_{\xi }}{\Lambda }\right) ^{2}\left( \frac{v_{\eta }}{\Lambda }%
\right) ^{8}\left( \frac{v_{\rho _{1}}}{\Lambda }\right) ^{4}\left(\frac{%
v_{\varphi }}{\Lambda }\right)^{2}m_{N_{n}},\hspace{1.5cm}  \notag \\
\left( M_{\nu }\right) _{22} &=&\sum_{n=1}^{2}\left( y_{2n}^{\left( V\right)
}\right) ^{2}f\left( m_{\func{Re}V^{0}},m_{\func{Im}V^{0}},m_{N_{n}}\right)
\left( \frac{v_{\xi }}{\Lambda }\right) ^{2}\left( \frac{v_{\eta }}{\Lambda }%
\right) ^{8}\left( \frac{v_{\rho _{1}}}{\Lambda }\right) ^{4}\left(\frac{%
v_{\varphi }}{\Lambda }\right)^{2}m_{N_{n}}=2\left( M_{\nu }\right) _{11},  \notag \\
\left( M_{\nu }\right) _{12} &=&\left( M_{\nu }\right)
_{21}=\sum_{n=1}^{2}\left( y_{1n}^{\left( V\right) }\right) \left(
y_{2n}^{\left( V\right) }\right) f\left( m_{\func{Re}V^{0}},m_{\func{Im}%
V^{0}},m_{N_{n}}\right) \left( \frac{v_{\xi }}{\Lambda }\right) ^{2}\left( 
\frac{v_{\eta }}{\Lambda }\right) ^{8}\left( \frac{v_{\rho _{1}}}{\Lambda }%
\right) ^{4}\left(\frac{v_{\varphi }}{\Lambda }\right)^{2}m_{N_{n}}=\left( M_{\nu }\right)
_{11}  \notag \\
\left( M_{\nu }\right) _{13} &=&\left( M_{\nu }\right) _{31}=\sqrt{2}%
\sum_{n=1}^{2}\left( y_{1n}^{\left( V\right) }\right) \left( y_{2n}^{\left(
V\right) }\right) f\left( m_{\func{Re}V^{0}},m_{\func{Im}V^{0}},m_{N_{n}}%
\right) \frac{v_{\xi }v_{\sigma }}{\Lambda ^{2}}\left( \frac{v_{\eta }}{%
\Lambda }\right) ^{8}\left( \frac{v_{\rho _{1}}}{\Lambda }\right) ^{4}\left(\frac{%
v_{\varphi }}{\Lambda }\right)^{2}m_{N_{n}}  \notag \\
\left( M_{\nu }\right) _{23} &=&\left( M_{\nu }\right) _{32}=\sqrt{2}%
\sum_{n=1}^{2}\left( y_{2n}^{\left( V\right) }\right) ^{2}f\left( m_{\func{Re%
}V^{0}},m_{\func{Im}V^{0}},m_{N_{n}}\right) \frac{v_{\xi }v_{\sigma }}{%
\Lambda ^{2}}\left( \frac{v_{\eta }}{\Lambda }\right) ^{8}\left( \frac{%
v_{\rho _{1}}}{\Lambda }\right) ^{4}\left(\frac{v_{\varphi }}{\Lambda }\right)^{2}m_{N_{n}} 
\notag \\
\left( M_{\nu }\right) _{33} &=&2\sum_{n=1}^{2}\left( y_{2n}^{\left(
V\right) }\right) ^{2}f\left( m_{\func{Re}V^{0}},m_{\func{Im}%
V^{0}},m_{N_{n}}\right) \left( \frac{v_{\sigma }}{\Lambda }\right)
^{2}\left( \frac{v_{\eta }}{\Lambda }\right) ^{8}\left( \frac{v_{\rho _{1}}}{%
\Lambda }\right) ^{4}\left(\frac{v_{\varphi }}{\Lambda }\right)^{2}m_{N_{n}}
\label{Mnuelements}
\end{eqnarray}
From the terms given above and the relations given by Eqs. (\ref{VEV}) and (%
\ref{Mnuelements}), we find that the charged lepton and light active
neutrino mass matrices are respectively given by: 
\begin{eqnarray}
M_{l} &=&\frac{v}{\sqrt{2}}\left( 
\begin{array}{ccc}
a_{1}\lambda ^{9} & 0 & a_{4}\lambda ^{4} \\ 
0 & a_{2}\lambda ^{5} & a_{5}\lambda ^{4} \\ 
0 & 0 & a_{3}\lambda ^{3}%
\end{array}%
\right) ,\hspace{1cm}\hspace{1cm} \\
M_{\nu } &=&\left( 
\begin{array}{ccc}
Z & Y & \sqrt{2}Y \\ 
Y & X & \sqrt{2}X \\ 
\sqrt{2}Y & \sqrt{2}X & 2X%
\end{array}%
\right),  \\
&\simeq &\sum_{n=1}^{2}\left( 
\begin{array}{ccc}
\left( y_{1n}^{\left( V\right) }\right) ^{2} & y_{1n}^{\left( V\right)
}y_{2n}^{\left( V\right) } & \sqrt{2}y_{1n}^{\left( V\right) }y_{2n}^{\left(
V\right) } \\ 
y_{1n}^{\left( V\right) }y_{2n}^{\left( V\right) } & \left( y_{2n}^{\left(
V\right) }\right) ^{2} & \sqrt{2}\left( y_{2n}^{\left( V\right) }\right) ^{2}
\\ 
\sqrt{2}y_{1n}^{\left( V\right) }y_{2n}^{\left( V\right) } & \sqrt{2}\left(
y_{2n}^{\left( V\right) }\right) ^{2} & 2\left( y_{2n}^{\left( V\right)
}\right) ^{2}%
\end{array}%
\right) \lambda ^{16}f\left( m_{\func{Re}V^{0}},m_{\func{Im}%
V^{0}},m_{N_{n}}\right) m_{N_{n}}
\end{eqnarray}%
where $a_{1}$, $a_{2}$, $b_{1}$, $c_{1}$, $y_{1n}^{\left( V\right) }$ and $%
y_{2n}^{\left( V\right) }$ are $\mathcal{O}(1)$ dimensionless quantities,
whereas $X$, $Y$, $Z$ are dimensionful parameters which are given by the
following relations: 
\begin{eqnarray}
X &\simeq &\sum_{n=1}^{2}\left( y_{2n}^{(V)}\right) ^{2}\lambda ^{16}f\left(
m_{\func{Re}V^{0}},m_{\func{Im}V^{0}},m_{N_{n}}\right) m_{N_{n}}, \\
Y &\simeq &\sum_{n=1}^{2}\left( y_{1n}^{(V)}\right) \left(
y_{2n}^{(V)}\right) \lambda ^{16}f\left( m_{\func{Re}V^{0}},m_{\func{Im}%
V^{0}},m_{N_{n}}\right) m_{N_{n}}, \\
Z &\simeq &\sum_{n=1}^{2}\left( y_{1n}^{(V)}\right) ^{2}\lambda ^{16}f\left(
m_{\func{Re}V^{0}},m_{\func{Im}V^{0}},m_{N_{n}}\right) m_{N_{n}},
\end{eqnarray}%
being $f\left( m_{\func{Re}V^{0}},m_{\func{Im}V^{0}},m_{N_{n}}\right) $ a
one loop function: 
\begin{eqnarray}
f\left( m_{\func{Re}V^{0}},m_{\func{Im}V^{0}},m_{N_{n}}\right) &=&\frac{1}{%
16\pi ^{2}}\left\{ \frac{\Lambda ^{2}}{m_{\func{Re}V^{0}}^{2}}-\frac{\Lambda
^{2}}{m_{\func{Im}V^{0}}^{2}}+\frac{m_{\func{Re}V^{0}}^{2}}{m_{\func{Re}%
V^{0}}^{2}-m_{N_{n}}^{2}}ln\left( \frac{m_{\func{Re}V^{0}}^{2}}{m_{N_{n}}^{2}%
}\right) -\frac{m_{\func{Im}V^{0}}^{2}}{m_{\func{Im}V^{0}}^{2}-m_{N_{n}}^{2}}%
ln\left( \frac{m_{\func{Im}V^{0}}^{2}}{m_{N_{n}}^{2}}\right) \right. \notag
\\
&&+\left. \left( \frac{m_{N_{n}}^{4}}{m_{\func{Re}V^{0}}^{2}(m_{\func{Re}%
V^{0}}^{2}-m_{N_{n}}^{2})}-\frac{m_{N_{n}}^{4}}{m_{\func{Im}V^{0}}^{2}(m_{%
\func{Im}V^{0}}^{2}-m_{N_{n}}^{2})}\right) ln\left( \frac{\Lambda
^{2}+m_{N_{n}}^{2}}{m_{N_{n}}^{2}}\right) \right\} .  
\end{eqnarray}
Let us note that the loop function has a quadratically divergent part, which arises from the $\frac{q_{\mu}q_{\nu}}{M_{V^{0}}^{2}}$ term of the heavy vector boson propagator.
\begin{table}[tbp]
\begin{center}
{\small 
\begin{tabular}{|c||c|c|c|c|}
\hline
\multirow{2}{*}{Observable} & \multirow{2}{*}{Model value} & 
\multicolumn{3}{|c|}{Experimental value} \\ \cline{3-5}
&  & $1\sigma $ range & $2\sigma $ range & $3\sigma $ range \\ \hline\hline
$m_{e}$ [MeV] & $0.487$ & $0.487$ & $0.487$ & $0.487$ \\ \hline
$m_{\mu }$ [MeV] & $102.8$ & $102.8\pm 0.0003$ & $102.8\pm 0.0006$ & $%
102.8\pm 0.0009$ \\ \hline
$m_{\tau }$ [GeV] & $1.75$ & $1.75\pm 0.0003$ & $1.75\pm 0.0006$ & $1.75\pm
0.0009$ \\ \hline
$m_{1}$ $[meV]$ & $49.19$ & $\cdots$ & $\cdots$ & $\cdots$ \\ \hline
$m_{2}$ $[meV]$ & $49.96$ & $\cdots$ & $\cdots$ & $\cdots$ \\ \hline
$m_{3}$ $[meV]$ & $0$ & $\cdots$ & $\cdots$ & $\cdots$ \\ \hline
$\Delta m_{21}^{2}$ [$10^{-5}$eV$^{2}$] (IH) & $7.55$ & $%
7.55_{-0.16}^{+0.20} $ & $7.20-7.94$ & $7.05-8.14$ \\ \hline
$\Delta m_{13}^{2}$ [$10^{-3}$eV$^{2}$] (IH) & $2.42$ & $%
2.42_{-0.04}^{+0.03} $ & $2.34-2.47$ & $2.31-2.51$ \\ \hline
$\delta $ [$^{\circ }$] (IH) & $309.719$ & $281_{-27}^{+23}$ & $229-328$ & $%
202-349$ \\ \hline
$\sin ^{2}\theta _{12}/10^{-1}$ (IH) & $3.20$ & $3.20_{-0.16}^{+0.20}$ & $%
2.89-3.59$ & $2.73-3.79$ \\ \hline
$\sin ^{2}\theta _{23}/10^{-1}$ (IH) & $5.33$ & $5.51_{-0.30}^{+0.18}$ & $%
4.91-5.84$ & $4.53-5.98$ \\ \hline
$\sin ^{2}\theta _{13}/10^{-2}$ (IH) & $2.248$ & $2.220_{-0.076}^{+0.074}$ & 
$2.07-2.36$ & $1.99-2.44$ \\ \hline
\end{tabular}%
}
\end{center}
\caption{Model and experimental values of the charged lepton masses,
neutrino mass squared splittings and leptonic mixing parameters for the
inverted (IH) mass hierarchy. The model values for CP violating phase and the light active neutrino masses are also shown. Notice that we have one massless light active neutrino state since there are two left handed handed Majorana neutrinos mediating the one loop level radiative seesaw mechanism that produces the light active neutrino masses. The experimental values of the charged lepton masses are taken from Ref.~\protect\cite{Bora:2012tx}, whereas the range for experimental
values of neutrino mass squared splittings and leptonic mixing parameters,
are taken from Ref.~\protect\cite{deSalas:2017kay}.}
\label{Tab}
\end{table}
The charged lepton masses, the neutrino mass squared splittings and the
leptonic mixing parameters can be very well reproduced for the scenario of
inverted neutrino mass ordering in terms of natural parameters of order one,
as shown in Table \ref{Tab}, starting from the following benchmark point: 
\begin{eqnarray}
a_{1} &\simeq &1.96168,\hspace{1cm}a_{2}\simeq 1.03698,\hspace{1cm}%
a_{3}\simeq 0.84294,\hspace{1.7cm}\left\vert a_{4}\right\vert \simeq 1.00752,%
\hspace{1cm}\arg \left( a_{4}\right) \simeq 218^{\circ },  \notag \\
a_{5} &\simeq &-0.597641,\hspace{0.5cm}X\simeq 16.5289~\mbox{meV},\hspace{%
0.3cm}Y\simeq -0.219701~\mbox{meV},\hspace{0.7cm}Z\simeq 49.5616~\mbox{meV},
\end{eqnarray}%
which corresponds to the eigenvalue problem solutions reproducing the
experimental values of the neutrino mass squared splittings and leptonic
mixing parameters. From the values of the $X$, $Y$ and $Z$ parameters given
above, it is possible to extract, using the light active neutrino mass
matrix elements, the mass of the left-handed sterile neutrinos ($m_{N}$) for
given values of neutrino--heavy-vector coupling and masses for the real and
imaginary parts of the neutral components of the heavy vector. As an
illustration, we fix the masses of the real and imaginary parts of the
neutral components of the heavy vector as $m_{\mathrm{Im}V^{0}}=1$ TeV, $m_{%
\mathrm{Re}V^{0}}=0.99$ TeV and we consider a neutrino--heavy-vector
coupling constant of order unity. For this case we found $m_{N}=245.811$ GeV.
Thus, the heavy vector will decay into an active neutrino and sterile
neutrino. Furthermore, the heavy vectors can be pair produced
at the LHC via vector boson fusion and Drell-Yan mechanism. In our
analysis we have set $\Lambda =3$ TeV, which is a typical value for the
cutoff scale in this kind of models with heavy resonances \cite%
{Zerwekh:2005wh,Barbieri:2009tx,Zerwekh:2009yu,Hernandez:2010iu,Hernandez:2010qp,Castillo-Felisola:2013jua,CarcamoHernandez:2017pei,Urbina:2018hnq}%
. A detailed study of the implications of our model at colliders and dark
matter goes beyond the scope of this paper and is deferred for a future
work. Studies of Heavy Majorana neutrino pair production at the LHC in the
framework of a very general anomaly free $U(1)_{X}$ model have been
performed in Refs. \cite{Das:2017flq,Das:2017deo}. In those $U(1)_{X}$
models, a linear seesaw mechanism at tree and one loop level can be
implemented to generate the masses for the light active neutrinos and the
dark matter relic density can be accommodated by having a scalar dark matter
candidate that annihilates into lepton-antilepton pair, where the exchange
of charged exotic leptons in the $t$ and $u$ channels mediates this
annihilation process, as done in Ref. \cite{Das:2017ski}.

In addition, we found a leptonic Dirac CP violating phase of $296.71^{\circ
} $ and a Jarlskog invariant close to about $-2.25\times 10^{-2}$ for the
inverted hierarchy. Now, let us consider the effective Majorana neutrino
mass parameter: 
\begin{equation}
m_{ee}=\left\vert \sum_{j}U_{ej}^{2}m_{\nu _{j}}\right\vert ,  \label{mee}
\end{equation}%
being $U_{ej}$ the elements of the PMNS matrix and $m_{\nu _{j}}$ the
Majorana neutrino masses. The neutrinoless double beta ($0\nu \beta \beta $)
decay amplitude is proportional to $m_{ee}$. With the model best fit values
in Table~\ref{Tab} we find 
\begin{equation}
m_{ee}\simeq 46.6813~\mbox{meV}\,.  \label{eff-mass-pred}
\end{equation}

\begin{figure}[]
\centering
\includegraphics[width=0.5\textwidth]{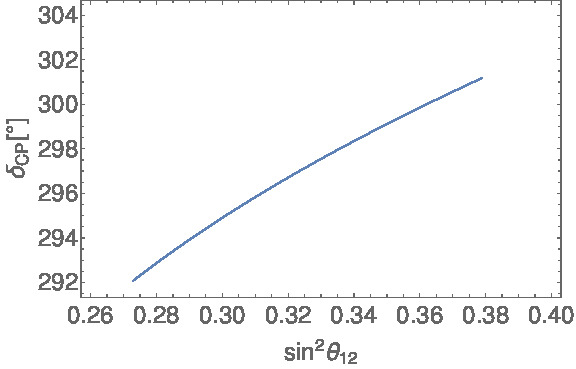}\newline
\caption{Correlation of the leptonic Dirac CP violating phase $\protect\delta%
_{CP}$ with the solar $\sin ^{2}\protect\theta _{12}$ mixing parameter.}
\label{CorrelationdeltaCP}
\end{figure}

\begin{figure}[]
\centering
\includegraphics[width=0.5\textwidth]{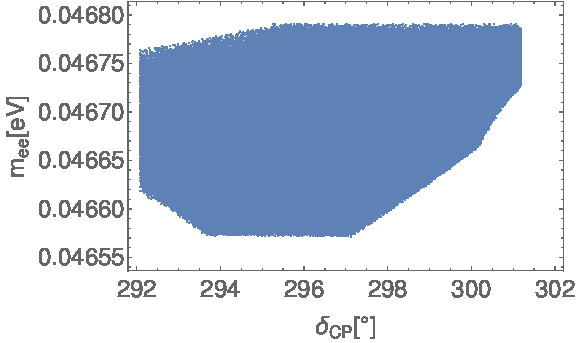}\newline
\caption{Correlation of the effective Majorana neutrino mass parameter $%
m_{ee}$ with leptonic Dirac CP violating phase $\protect\delta_{CP}$.}
\label{Correlationmee}
\end{figure}

Figures~\ref{CorrelationdeltaCP} and \ref{Correlationmee} show the
correlations of the leptonic Dirac CP violating phase $\delta_{CP}$ with the
solar $\sin ^{2}\theta _{12}$ mixing parameter and the effective Majorana
neutrino mass parameter $m_{ee}$ vs the leptonic Dirac CP violating phase $%
\delta_{CP}$. To obtain these Figures, we randomly varied the lepton sector model parameters around their best fit values in such a way that the resulting neutrino mass squared splittings, leptonic mixing parameters and leptonic Dirac CP violating phase be inside the $3\sigma $ experimentally allowed range.

As indicated by Figures~\ref{CorrelationdeltaCP} and \ref%
{Correlationmee}, our model predicts solar mixing parameter, leptonic Dirac
CP violating phase and effective Majorana neutrino mass parameter in the
ranges $0.27\lesssim\sin ^{2}\theta _{12}\lesssim 0.38$, $%
292^{\circ}\lesssim\delta\lesssim 301^{\circ}$ and $0.04655$ eV$\lesssim
m_{ee}\lesssim$ $0.04679$ eV, respectively.

The range of values given above for the effective Majorana neutrino mass
parameter $m_{ee}$ are within the declared reach of the next-generation
bolometric CUORE experiment \cite{Alessandria:2011rc} or, more
realistically, of the next-to-next-generation ton-scale $0\nu \beta \beta $%
-decay experiments. The current most stringent experimental upper limit $%
m_{ee}\leq 160$ meV is set by $T_{1/2}^{0\nu \beta \beta }(^{136}\mathrm{Xe}%
)\geq 1.1\times 10^{26}$ yr at 90\% C.L. from the KamLAND-Zen experiment 
\cite{KamLAND-Zen:2016pfg}.

\section{An alternative model.}
\label{alter}
Given that the normal hierarchy is favored over more than $3\sigma $ than
the inverted one, in this section we propose an alternative model based on
the $S_{3}\times Z_{3}\times Z_{4}\times Z_{5}\times Z_{8}$ discrete group,
where the assignments of the leptonic and scalar fields under such discrete
group are shown in Tables \ref{leptonsalternativemodel} and \ref{scalarsalternativemodel}, respectively. In addition, the $S_{3}\times Z_{3}\times
Z_{4}\times Z_{5}\times Z_{8}$  assignment of the heavy vector in the
fundamental $SU\left( 2\right) _{L}$ representation is:%
\begin{equation}
V_{\mu }\sim \left( \mathbf{1},0,2,0,0\right) .
\end{equation}
We consider the following VEV configurations for the $S_3$ scalar doublets, which are natural solutions of
the scalar potential minimization conditions:
%We assume that the $S_{3}$ scalar doublets have the following VEV patterns:
\begin{equation}
\left\langle \chi \right\rangle =v_{\chi }\left( 1,1\right) ,\hspace{1cm}%
\hspace{1cm}\left\langle \eta \right\rangle =v_{\chi }\left( 1,-1\right) ,%
\hspace{1cm}\hspace{1cm}\left\langle \xi \right\rangle =v_{\xi }\left( 1,%
\sqrt{2}\right).
\end{equation}
With the previously specified particle content, we get the following lepton
Yukawa terms as well as the interaction terms of the heavy vector with the heavy left handed Majorana
neutrinos:
\begin{eqnarray}
\tciLaplace _{Y}^{\left( l\right) } &=&y_{11}^{\left( l\right) }\overline{l}%
_{1L}\phi l_{1R}\frac{\varphi \Delta ^{4}\sigma ^{3}\rho }{\Lambda ^{9}}%
+y_{21}^{\left( l\right) }\overline{l}_{L}\phi l_{1R}\frac{\eta \Delta ^{4}}{%
\Lambda ^{9}}\left[ \sigma ^{3}\rho +x_{1}\left( \sigma ^{\ast }\right)
^{2}\left( \rho ^{\ast }\right) ^{2}\right]\notag \\
&&+y_{12}^{\left( l\right) }\overline{l}_{1L}\phi l_{2R}\frac{\left( \chi
\xi ^{\ast }\xi ^{\ast }\right) _{\mathbf{\mathbf{1}^{\prime }}}\sigma
^{2}\rho }{\Lambda ^{6}}+y_{22}^{\left( l\right) }\overline{l}_{L}\phi l_{2R}%
\frac{\chi }{\Lambda ^{9}}\left[ \sigma ^{3}\rho +x_{2}\left( \sigma ^{\ast
}\right) ^{2}\left( \rho ^{\ast }\right) ^{2}\right]\notag \\
&&+y_{13}^{\left( l\right) }\overline{l}_{1L}\phi l_{3R}\frac{\sigma ^{3}}{%
\Lambda ^{3}}+y_{23}^{\left( l\right) }\overline{l}_{L}\phi l_{3R}\frac{\chi
\sigma ^{3}}{\Lambda ^{4}}+y_{33}^{\left( l\right) }\overline{l}_{L}\phi
l_{3R}\frac{\chi \left( \sigma ^{\ast }\right) ^{2}}{\Lambda ^{4}}\notag \\
&&+\frac{1}{2}\sum_{n=1}^{2}m_{N_{n}}\overline{N}_{nL}N_{nL}^{C}+h.c,
\end{eqnarray}
\begin{equation}
-\mathcal{L}_{VlN}=\sum_{n=1}^{2}y_{1n}^{\left( V\right) }\overline{l}%
_{1L}\gamma ^{\mu }V_{\mu }N_{nL}\frac{\xi \Delta ^{4}\sigma ^{2}\rho }{%
\Lambda ^{8}}+\sum_{n=1}^{2}y_{2n}^{\left( V\right) }\overline{l}_{L}\gamma
^{\mu }V_{\mu }N_{nL}\frac{\Phi \Delta ^{4}\sigma ^{2}\rho }{\Lambda ^{8}}.
\end{equation}
\begin{table}[tbp]
\centering%
\begin{tabular}{|c|c|c|c|c|c|c|c|}
\hline
& $l_{1L}$ & $l_{L}$ & $l_{1R}$ & $l_{2R}$ & $l_{3R}$ & $N_{1L}$ & $N_{2L}$
\\ \hline
$S_{3}$ & $\mathbf{1}$ & $\mathbf{2}$ & $1^{\prime }$ & $1^{\prime }$ & $%
\mathbf{1^{\prime }}$ & $\mathbf{1}$ & $\mathbf{1}$ \\ \hline
$Z_{3}$ & $-1$ & $-1$ & $0$ & $0$ & $-1$ & $0$ & $0$ \\ \hline
$Z_{4}$ & $0$ & $0$ & $1$ & $0$ & $0$ & $0$ & $0$ \\ \hline
$Z_{5}$ & $1$ & $1$ & $4$ & $4$ & $4$ & $0$ & $0$ \\ \hline
$Z_{8}$ & $0$ & $0$ & $4$ & $0$ & $0$ & $4$ & $4$ \\ \hline
\end{tabular}%
\caption{$S_{3}\times Z_{3}\times Z_{4}\times Z_{5}\times Z_{6}$ assignments of the leptonic fields in the alternative model.}
\label{leptonsalternativemodel}
\end{table}
\begin{table}[tbp]
\centering%
\begin{tabular}{|c|c|c|c|c|c|c|c|c|}
\hline
& $\phi $ & $\chi $ & $\eta $ & $\xi $ & $\sigma $ & $\varphi $ & $\rho $ & $%
\Delta $ \\ \hline
$S_{3}$ & $\mathbf{1}$ & $\mathbf{2}$ & $\mathbf{2}$ & $2$ & $1^{\prime }$ & 
$\mathbf{1}$ & $\mathbf{1}$ & $1$ \\ \hline
$Z_{3}$ & $0$ & $0$ & $0$ & $0$ & $0$ & $0$ & $-1$ & $0$ \\ \hline
$Z_{4}$ & $0$ & $0$ & $-1$ & $-2$ & $0$ & $-1$ & $0$ & $0$ \\ \hline
$Z_{5}$ & $0$ & $0$ & $0$ & $-2$ & $-1$ & $0$ & $0$ & $0$ \\ \hline
$Z_{8}$ & $0$ & $0$ & $0$ & $0$ & $0$ & $0$ & $0$ & $-1$ \\ \hline
\end{tabular}%
\caption{$S_{3}\times Z_{3}\times Z_{4}\times Z_{5}\times Z_{6}$ assignments of the scalar fields.}
\label{scalarsalternativemodel}
\end{table}
From the terms given above and the relations given by Eqs. (\ref{VEV}) and (%
\ref{Mnuelements}), we find that the charged lepton and light active
neutrino mass matrices are respectively given by: 
\begin{eqnarray}
M_{l} &=&\left( 
\begin{array}{ccc}
a_{11}\lambda ^{9} & a_{12}\lambda ^{6} & a_{13}\lambda ^{3} \\ 
a_{21}\lambda ^{9} & a_{22}\lambda ^{5} & a_{23}\lambda ^{4} \\ 
a_{31}\lambda ^{9} & a_{32}\lambda ^{5} & a_{33}\lambda ^{3} \\ 
&  & 
\end{array}%
\right) \frac{v}{\sqrt{2}},\hspace{1cm}\hspace{1cm} \\
M_{\nu } &=&\left( 
\begin{array}{ccc}
Z & Y & \sqrt{2}Y \\ 
Y & X & \sqrt{2}X \\ 
\sqrt{2}Y & \sqrt{2}X & 2X%
\end{array}%
\right) ,  \notag \\
&\simeq &\sum_{n=1}^{2}\left( 
\begin{array}{ccc}
\left( y_{1n}^{\left( V\right) }\right) ^{2} & y_{1n}^{\left( V\right)
}y_{2n}^{\left( V\right) } & \sqrt{2}y_{1n}^{\left( V\right) }y_{2n}^{\left(
V\right) } \\ 
y_{1n}^{\left( V\right) }y_{2n}^{\left( V\right) } & \left( y_{2n}^{\left(
V\right) }\right) ^{2} & \sqrt{2}\left( y_{2n}^{\left( V\right) }\right) ^{2}
\\ 
\sqrt{2}y_{1n}^{\left( V\right) }y_{2n}^{\left( V\right) } & \sqrt{2}\left(
y_{2n}^{\left( V\right) }\right) ^{2} & 2\left( y_{2n}^{\left( V\right)
}\right) ^{2}%
\end{array}%
\right) \lambda ^{16}f\left( m_{\func{Re}V^{0}},m_{\func{Im}%
V^{0}},m_{N_{n}}\right) m_{N_{n}},
\end{eqnarray}%
where $a_{11}$, $a_{12}$, $a_{13}$, $a_{21}$, $a_{22}$, $a_{23}$, $a_{31}$, $%
a_{32}$, $a_{33}$, $y_{1n}^{\left( V\right) }$ and $y_{2n}^{\left( V\right) }
$ are $\mathcal{O}(1)$ dimensionless quantities, whereas $X$, $Y$, $Z$ are
dimensionful parameters which are given by the following relations:%
\begin{eqnarray}
X &\simeq &\sum_{n=1}^{2}\left( y_{2n}^{(V)}\right) ^{2}\lambda ^{16}f\left(
m_{\func{Re}V^{0}},m_{\func{Im}V^{0}},m_{N_{n}}\right) m_{N_{n}}, \\
Y &\simeq &\sum_{n=1}^{2}\left( y_{1n}^{(V)}\right) \left(
y_{2n}^{(V)}\right) \lambda ^{16}f\left( m_{\func{Re}V^{0}},m_{\func{Im}%
V^{0}},m_{N_{n}}\right) m_{N_{n}}, \\
Z &\simeq &\sum_{n=1}^{2}\left( y_{1n}^{(V)}\right) ^{2}\lambda ^{16}f\left(
m_{\func{Re}V^{0}},m_{\func{Im}V^{0}},m_{N_{n}}\right) m_{N_{n}}.
\end{eqnarray}
The charged lepton masses, the neutrino mass squared splittings and the
leptonic mixing parameters can be very well reproduced for the scenario of
normal neutrino mass ordering in terms of natural parameters of order one,
as shown in Table \ref{Tab:leptonObsNH}, starting from the following benchmark point: 
\begin{eqnarray}
a_{11} &=&-1.04606 + 0.07753i,\hspace{0.8cm}a_{12}=-0.942761 - 0.154021i,%
\hspace{0.8cm}a_{13}=0.754455-0.0433344i,\notag \\
a_{21} &=&-1.09641 + 0.0882596i,\hspace{0.8cm}a_{22}=0.807509 - 0.0359955 i,%
\hspace{0.8cm}a_{23}=-0.782795 - 0.125467i,\notag \\
a_{31} &=&1.1238 + 0.124818 i,\hspace{0.8cm}a_{32}=0.59038 + 0.0469802 i,\hspace{%
0.8cm}a_{33}=0.490204 + 0.0368452i,\notag\\
%\begin{equation}
X &=& 13.9813\text{meV},\hspace{1cm}Y=9.73655\text{meV},\hspace{1cm}Z=17.2485%
\text{meV},
\end{eqnarray}
%\end{equation}
which corresponds to the eigenvalue problem solutions reproducing the experimental values of the physical observables of the lepton sector: charged lepton masses, neutrino mass squared
differences and leptonic mixing parameters. In addition the values of the $X$, $Y$ and $Z$ parameters given above can be reproduced for the following masses:
\begin{equation}
m_{\mathrm{Im}V^{0}}=1.5\text{TeV},\hspace{1cm}m_{\mathrm{Re}V^{0}}=1.6\text{TeV},\hspace{1cm}m_N=159\text{MeV}.    
\end{equation}
As we will see in the next section, the masses of the neutral components of the heavy vector as well as of the left handed Majorana neutrinos given above, are inside the allowed region of parameter space consistent with the constraints arising for charged lepton flavor violation. In addition, the rates for the charged lepton flavor violating decays obtained for those mass values are within the reach of the forthcoming experiments.
\begin{table}[H]
\centering
%\captionsetup{width=0.6\textwidth}
\begin{tabular}{|c|c|c|c|c|}
\hline
\multirow{2}{*}{Observable} & \multirow{2}{*}{Model Value} & 
\multicolumn{3}{c|}{Experimental value} \\ \cline{3-5}
&  & $1\sigma$ range & $2\sigma$ range & $3\sigma$ range \\ \hline\hline
$m_{e}$ [MeV] & $0.487$ & $0.487$ & $0.487$ & $0.487$ \\ \hline
$m_{\mu }$ [MeV] & $102.8$ & $102.8\pm 0.0003$ & $102.8\pm 0.0006$ & $%
102.8\pm 0.0009$ \\ \hline
$m_{\tau }$ [GeV] & $1.75$ & $1.75\pm 0.0003$ & $1.75\pm 0.0006$ & $1.75\pm
0.0009$ \\ \hline
$m_{1}$ $[meV]$ & $0$ & $\cdots$ & $\cdots$ & $\cdots$ \\ \hline
$m_{2}$ $[meV]$ & $8.67$ & $\cdots$ & $\cdots$ & $\cdots$ \\ \hline
$m_{3}$ $[meV]$ & $50$ & $\cdots$ & $\cdots$ & $\cdots$ \\ \hline
$\Delta m_{21}^2$ $[10^{-5}\,eV^2]$ & $7.55$ & $7.55^{+0.20}_{-0.16}$ & $7.20 -
7.94$ & $7.05 - 8.14$ \\ \hline
$\Delta m_{31}^2$ $[10^{-3}\,eV^2]$ & $2.50$ & $2.50\pm0.03$ & $2.44 - 2.57$ & $%
2.41-2.60$ \\ \hline
$\sin^2(\theta_{12})/10^{-1}$ & $3.20$ & $3.20^{+0.20}_{-0.16}$ & $%
2.89-3.59$ & $2.73-3.79$ \\ \hline
$\sin^2(\theta_{23})/10^{-1}$ & $5.47$ & $5.47^{+0.20}_{-0.30}$ & $%
4.67-5.83$ & $4.45-5.99$ \\ \hline
$\sin^2(\theta_{13})/10^{-2}$ & $2.160$ & $2.160^{+0.083}_{-0.069}$ & $%
2.03-2.34$ & $1.96-2.41$ \\ \hline
$\delta_{CP}$ & $218^\circ$ & $218^{+38^\circ}_{-27^\circ}$ & $182^\circ-315^\circ$ & $157^\circ-349^\circ$ \\
\hline
\end{tabular}
\caption{Model and experimental values of the charged lepton masses,
neutrino mass squared splittings and leptonic mixing parameters for the
normal (NH) mass hierarchy. The model values for CP violating phase and the light active neutrino masses are also shown. Notice that we have one massless light active neutrino state since there are two left handed handed Majorana neutrinos mediating the one loop level radiative seesaw mechanism that produces the light active neutrino masses. The experimental values of the charged lepton masses are taken
from Ref.~\protect\cite{Bora:2012tx}, whereas the range for experimental
values of neutrino mass squared splittings and leptonic mixing parameters,
are taken from Ref.~\protect\cite{deSalas:2017kay}.}
\label{Tab:leptonObsNH}
\end{table}
\section{Charged lepton flavor violating decay constraints.}
\label{LFV} 
\begin{figure}[tbp]
\includegraphics[width=0.49\textwidth]{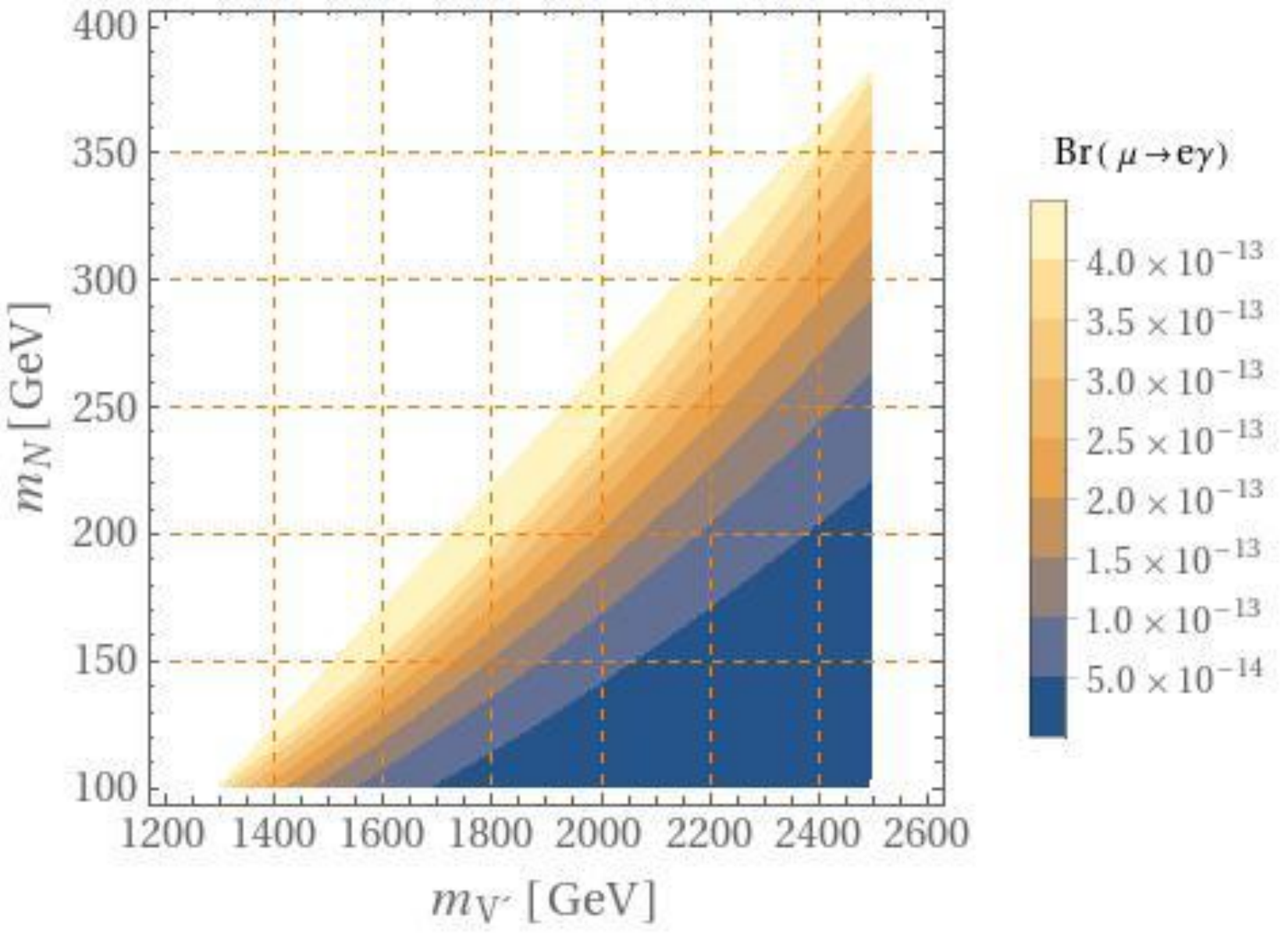}
\caption{Allowed parameter space in the $m_{V^{\pm }}-m_{N}$ plane
consistent with the LFV constraints.}
\label{LFVplot}
\end{figure}

In this section we will determine the constraints that the
charged lepton flavor violating (CLFV) decays $\mu \rightarrow e\gamma $, $%
\tau \rightarrow \mu \gamma $ and $\tau \rightarrow e\gamma $ imposed in the
parameter space of our model. These CLFV processes appear at one loop level
and are mediated by sterile neutrinos together with the charged components
of the heavy vector doublet. The Branching ratio of the $l_{i}\rightarrow
l_{j}\gamma $ decay is given by: \cite%
{Ilakovac:1994kj,Deppisch:2004fa,Lindner:2016bgg}: 
\begin{eqnarray}
Br\left( l_{i}\rightarrow l_{j}\gamma \right) &=&\frac{\alpha
_{W}^{3}s_{W}^{2}m_{l_{i}}^{5}}{256\pi ^{2}m_{V^{\pm }}^{4}\Gamma _{i}}%
\left\vert\sum_{n=1}^{2}G\left( \frac{m_{N_{n}}^{2}}{m_{V^{\pm }}^{2}}\right)\right\vert ^{2},%
\hspace{0.5cm}\hspace{0.5cm}\hspace{0.5cm},  \notag \\
G\left( x\right) &=&-\frac{2x^{3}+5x^{2}-x}{4\left( 1-x\right) ^{2}}-\frac{%
3x^{3}}{2\left( 1-x\right) ^{4}}\ln x,
\label{Brmutoegamma}
\end{eqnarray}
where $m_{V^{\pm }}$ are the masses of the charged components of the $SU_{2L}$ vector doublet $V_{\mu}$, whereas $m_{N_{n}}$ ($n=1,2$)
corresponds to the masses of the left handed Majorana neutrinos $N_{nL}$. To simplify our analysis, we chose a benchmark scenario where the left handed Majorana neutrinos $N_{nL}$ are degenerate with mass equal to $%
m_{N}$. Figure \ref{LFV} shows the allowed parameter space in the $m_{V^{\pm
}}-m_{N} $ plane consistent with the constraints arising from charged lepton
flavor violating decays. As seen from Figure \ref{LFVplot}, the obtained
values for the branching ratio of $\mu \rightarrow e\gamma $ decay are below
its experimental upper limit of $4.2\times 10^{-13}$ since these values are
located in the range $5\times 10^{-14}\lesssim Br\left( \mu \rightarrow
e\gamma \right) \lesssim 2.5\times 10^{-13}$, for charged heavy vector and
heavy Majorana neutrino masses in the ranges $1.3$ TeV$\lesssim m_{V^{\pm
}}\lesssim 2.5$ TeV and $100$ GeV$\lesssim m_{N}\lesssim 380$ GeV,
respectively. It is worth mentioning that we have considered heavy vectors lighter than about $2.5$ TeV in order to comply with the constraints arising from perturbative unitarity \cite{Saez:2018off}. In the same region of parameter space, the obtained branching ratios for the $\tau \rightarrow \mu \gamma $ and $\tau \rightarrow e\gamma $
decays can reach values of the order of $10^{-10}$, which is below their
corresponding upper experimental bounds of $4.4\times 10^{-8}$ and $%
3.3\times 10^{-8}$, respectively. Consequently, our model predicts charged
lepton flavor violating processes within the reach of the forthcoming
experiments. 

\section{Conclusions}

\label{conclusions} We have built two viable extensions of the
SM with a heavy vector in the fundamental $SU\left( 2\right) _{L}$
representation and several gauge singlet scalars that successfully explain the
observed fermion mass spectrum and mixing parameters at low energies. The
models incorporate the $S_{3}$ family symmetry, which is supplemented by other auxiliary cyclic symmetries. The
observed hierarchical structure of the SM charged fermion mass spectrum and
mixing parameters comes from the spontaneous breaking of both the $S_{3}$ and the different cyclic symmetries. The masses for
the light active neutrinos are generated via a one loop level radiative
seesaw mechanism. In the first model, the obtained quark mass spectrum, CKM parameters, CP
violating phase are compatible with their experimental values, whereas the
resulting leptonic mixing parameters agree with current neutrino data only
in the case of inverted ordering, featuring in this scenario an excellent
agreement. The second model allows to accommodate the experimental values for the physical observables of the lepton sector in the scenario of normal neutrino mass hierarchy. In addition, the proposed models are consistent with the constraints arising from charged lepton flavor violating processes, which restrict the charged heavy vector mass to be larger than about $1.3$ TeV and the heavy Majorana neutrino masses in the range $100$ GeV$\lesssim m_{N}\lesssim 380$ GeV. We have found that our model predicts charged lepton flavor violating decays within the reach of the current sensitivity of the forthcoming charged lepton flavor violation experiments%For the
%Furthermore, an effective Majorana neutrino mass parameter of $%m_{ee}\simeq 46.7$ meV is predicted by our model in the scenario of inverted neutrino mass spectrum.7

\section*{Acknowledgments}
This research has received funding from Fondecyt (Chile), Grants No.~1170803
and No.1160423 , CONICYT PIA/Basal FB0821 and PIA/ACT-1406. J.V.
acknowledges PIIC/DGIP for supporting this work.
\appendix

\section{The product rules for $S_{3}$.}
\label{S3}
The $S_{3}$ discrete group contains 3 irreducible representations: $\mathbf{1%
}$, $\mathbf{1}^{\prime }$ and $\mathbf{2}$. Considering $\left(
x_{1},x_{2}\right) ^{T}$\ and $\left( y_{1},y_{2}\right) ^{T}$ as the basis
vectors for two $S_{3}$ doublets and $y%
%TCIMACRO{\U{b4}}%
%BeginExpansion
{\acute{}}%
%EndExpansion
$ an $S_{3}$ non trivial singlet, the multiplication rules of the $S_{3}$
group for the case of real representations take the form \cite%
{Ishimori:2010au}: 
\begin{equation}
\left( 
\begin{array}{c}
x_{1} \\ 
x_{2}%
\end{array}%
\right) _{\mathbf{2}}\otimes \left( 
\begin{array}{c}
y_{1} \\ 
y_{2}%
\end{array}%
\right) _{\mathbf{2}}=\left( x_{1}y_{1}+x_{2}y_{2}\right) _{\mathbf{1}%
}+\left( x_{1}y_{2}-x_{2}y_{1}\right) _{\mathbf{1}^{\prime }}+\left( 
\begin{array}{c}
x_{2}y_{2}-x_{1}y_{1} \\ 
x_{1}y_{2}+x_{2}y_{1}%
\end{array}%
\right) _{\mathbf{2}},  \label{6}
\end{equation}%
\begin{equation}
\left( 
\begin{array}{c}
x_{1} \\ 
x_{2}%
\end{array}%
\right) _{\mathbf{2}}\otimes \left( y%
%TCIMACRO{\U{b4}}%
%BeginExpansion
{\acute{}}%
%EndExpansion
\right) _{\mathbf{1}^{\prime }}=\left( 
\begin{array}{c}
-x_{2}y%
%TCIMACRO{\U{b4} }%
%BeginExpansion
{\acute{}}
%EndExpansion
\\ 
x_{1}y%
%TCIMACRO{\U{b4}}%
%BeginExpansion
{\acute{}}%
%EndExpansion
\end{array}%
\right) _{\mathbf{2}},\hspace{1cm}\hspace{1cm}\left( x%
%TCIMACRO{\U{b4}}%
%BeginExpansion
{\acute{}}%
%EndExpansion
\right) _{\mathbf{1}^{\prime }}\otimes \left( y%
%TCIMACRO{\U{b4}}%
%BeginExpansion
{\acute{}}%
%EndExpansion
\right) _{\mathbf{1}^{\prime }}=\left( x%
%TCIMACRO{\U{b4}}%
%BeginExpansion
{\acute{}}%
%EndExpansion
y%
%TCIMACRO{\U{b4}}%
%BeginExpansion
{\acute{}}%
%EndExpansion
\right) _{\mathbf{1}}.  \label{7}
\end{equation}

\end{document}